\documentclass[11pt,twoside]{article}

\usepackage{asp2006}
\usepackage{epsf}
\usepackage{lscape}

\begin{document}
\title{Interferometric Observations of Algol}   
\author{Tam\'as Borkovits,\altaffilmark{1} Szil\'ard Csizmadia,\altaffilmark{2,3}  
Zsolt Paragi,\altaffilmark{4,3}, L\'aszl\'o Sturmann,\altaffilmark{5} Judit Sturmann,\altaffilmark{5} 
Chris Farrington,\altaffilmark{5} Harold A. McAlister,\altaffilmark{5} Theo A. ten Brummelaar,\altaffilmark{5} Nils H. Turner\altaffilmark{5}}   
\altaffiltext{1}{Baja Astronomical Observatory, H-6500 Baja, Szegedi \'ut, Kt. 766, Hungary}
\altaffiltext{2}{Institute of Planetary Research, DLR Rutherfordstr. 2. D-12489, Berlin, Germany }    
\altaffiltext{3}{MTA Research Group for Physical Geodesy and Geodynamics, H-1585 Budapest, P. O. Box 585., Hungary}
\altaffiltext{4}{Joint Institute for VLBI in Europe, Postbus 2, 7990\,AA Dwingeloo, The Netherlands}
\altaffiltext{5}{Center for High Angular Resolution Astronomy, Georgia State University,P. O. Box 3969, Atlanta, GA 30302}

\begin{abstract} We determined the spatial orientation of the Algol AB close pair
orbital plane using optical interferometry with the CHARA Array, and radio 
interferometry with the European VLBI Network (EVN). We found the longitude of the line of nodes for the close pair being
$\Omega_1=48\deg\pm2\deg$ and the mutual inclination of 
the orbital planes of the close and the wide pairs being $95\deg\pm3\deg$.
This latter value differs by $5\deg$ from the formerly known $100\deg$
which would imply a very fast inclination variation of the system, not supported by the photometric observations. We also investigated the
dynamics of the system with numerical integration of the equations of
motions using our result as an initial condition. We found large variations 
in the inclination of the close pair (its amplitude $\sim 170\deg$)
with a period of about 20 millenia. This result is in good agreement
with the photometrically observed change of amplitude in Algol's
primary minimum.
\end{abstract}


\section{Introduction}

Algol consists of a semi-detached eclipsing binary with an orbital
period of 2.87 days (B8V + K2IV) with an F1IV spectral type star
revolving around the binary every 680 days. The third
component was succesfully observed by speckle interferometry, 
and its orbit was precisely determined by \citet{bonneau79}. This result
was refined by using the Mark III optical stellar interferometer \citep*{panetal93}. 

In the radio regime, \citet{lestradeetal93} detected positional
displacement during the orbital revolution of the AB pair using the VLBI
technique, and identified the K-subgiant as the source of radio
emission. From their measurements \citet*{kiselevaetal98} calculated
$i_\mathrm{m}=100\deg$ for the mutual inclination of the close and wide orbital planes.
This value has been widely accepted since then. However, this mutual
inclination value cannot be correct due to dynamical considerations, because
it would produce a fast variation in the observable inclination of the eclipsing
subsystem, resulting in a fast eclipse depth variation which contradicts to
the more than century-long photometric observations \citep{soderhjelm80}.

The aim of this study was to constrain the mutual inclination of the
system better, requiring the measurement of the longitude of the node
for the close pair. The other orbital elements are well-known from
spectroscopic or photometric data, but there is a controversy in the
value of $\Omega_1$. Because the expected apparent size of the
close binary semi-major axis is of the order of 2 milliarcseconds
(mas), we carried out both optical and radio interferometry measurements.

\section{Observations and results}

We carried out interferometric observations both in the optical/near-infrared, and 
in the radio regime.
The optical interferometry observations were done by the CHARA Array on three nights (2, 3 and 4 December, 2006) in the
$K_s$ band. The radio measurements were carried out with a subset of the European VLBI Network (EVN) 
on 14-15 December 2006 at 5~GHz.
At this observation the e-VLBI technique was applied, where  the
telescopes stream the data to the central data processor (JIVE, Dwingeloo,
the Netherlands) instead of recording. 
The participating telescopes were Cambridge and 
Jodrell Bank (UK),  Medicina (Italy), Onsala (Sweden), Toru\'n (Poland) and 
the Westerbork phased array (the Netherlands). 
The measurements and the data reductions were carried out according to the standard
procedures. (Details can be read in \citealp{csizmadiaetal09}.)

For the analysis of the CHARA visibilities we developed a model which was very close to the one of
\citet{wilsondevinney71} which is based on the Roche-model, but it was implemented
in IDL to restore the surface intensities into a matrix and to calculate the sky-projected
picture of the system. Since we had only about two dozen visibility measurements, we wanted to
limit the number of free parameters, choosing just three: the angular
size of the semi-major axis $a_1$, the surface brightness ratio in $K_{s}$ band $J_{K_s}$, and the angle $\Omega_1$. 
All other parameters were fixed
according to the values given in \citet{wilsonetal72} or in \citet{kim89}. 
After an extended grid search we got $a_1=2.28\pm0.02\mathrm{mas}$, $J_{K_s}=0.330\pm0.01$, 
and $\Omega_1=48\deg\pm2\deg$, respectively.

The VLBI data were gained during a secondary minimum, which, according to our simultaneous
optical photometry, occurred at $t_0=2\,454\,084.360\pm0.003$ \citep{biroetal07}. 
The measurements lasted 9 hours, i.e. $\approx13\%$ of the orbital period, but as the projected
movement of the source is maximal around the minima, the covered projected orbital arc
is significantly larger. Due to the large inclination, this projected
arc is almost a straight line parallel to the nodal line, which makes the determination
of the longitude of the node more easier. Unfortunately, during our measurements
a radio flare occured which significantly reduced the accuracy of our position determination.
Our best fit gave $\Omega_1=52\deg\pm3\deg$, but we consider this result
less reliable because it is difficult to assess how much the radio 
flare as well as atmospheric phase fluctuations might have affected 
our data (e.g. the displacement of the source during our observation 
was almost twice of the calculated from Keplerian revolution).

Due to the above mentioned facts, here we concentrate mainly on the CHARA results. 
The true size of the obtained semi-major axis and surface brightness ratio are in 
very good agreement with previous results, which makes our nodal result also plausible.
According to the CHARA results, the longitude of the node is $\Omega_1=48\deg\pm2\deg$ 
with an ambiguity of $180\deg$. With the VLBI measurements we can resolve the
$\pm180\deg$ ambiguity and conclude that $\Omega_1 = 48\deg\pm2\deg$.
This is in excellent agreement with the value determined from
polarimetric measurements ($\Omega_1 = 47\deg\pm7\deg$, \citealp{rudy79}),
indicating that polarimetry is an efficient tool to determine the
spatial orientation of the orbits.

At this point we can determine the mutual inclination $i_m$, and we get $i_m=95\deg\pm3\deg$. 
This value is, however, closer to the exact perpendicularity than the $100\deg$  
which was based on the measurements of \citet{lestradeetal93}.

\begin{figure}
\plotone{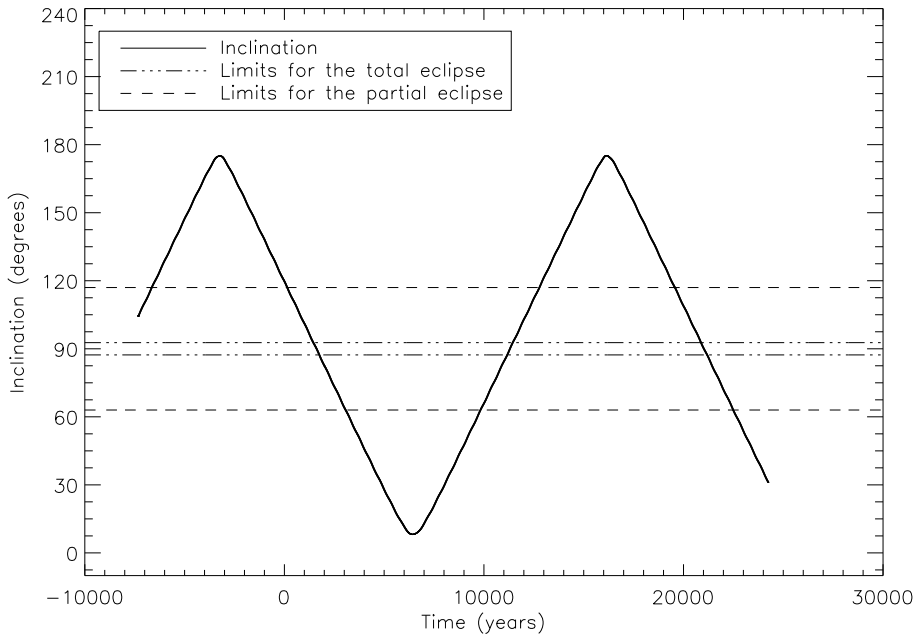}
\caption{The variation of the inclination of Algol AB with time. Solid line
represents the inclinitaion of the close pair observable from the Earth.
Between the dashed lines the system shows partial eclipses, moreover between
the dashed-dotted lines the system shows total eclipses for a short time.
 \label{fig:Algol_long}}
\end{figure}

\section{Dynamics of the system}

In order to investigate the dynamical behaviour of Algol in the near past and
future, we carried out numerical integration of the orbits for the triple 
system. Detailed description of our code can be found in \citet*{borkovitsetal04}. This
code simultaneously integrates the equations of the orbital motions and the
Eulerian equations of stellar rotation. The code also includes stellar dissipation,
but the short time interval of the data allows that term to be ignored. Our input 
parameters for Algol AB were
taken from \citet{kim89} with the
exception of $\Omega_1$ which was set to $48\deg$ in accordance with 
our CHARA result. The orbital elements of the wide orbit were almost identical to
the elements given by \citet{panetal93}.
As further input parameters, the $k_2$, $k_3$ internal structure constants for
the binary members were taken from the tables of \citet{claretgimenez92}. 

Here we consider only the observable inclination of the eclipsing pair. 
The computed variation of inclination between 7500 BC
and 22\,500 AD was plotted in Fig.~\ref{fig:Algol_long}.
Note that Algol AB does not show eclipses when the inclination is lower than 
$63\deg$ or higher than $117\deg$. It shows partial eclipses if the 
inclination is between $63\deg$--$117\deg$ and moreover, it shows total eclipses when the
inclination $87.3\deg$--$92.7$. 
As one can see, Algol started to show eclipses shortly after the beginning of the
Christian calendar. Total eclipses occured during an almost 300-year-long interval
at the end of the medieval, and at the dawn of the modern, scientific era. 
According to our integrations in the time of the discovery
as a variable star \citep{montanari671}, the inclination of the close pair was about
$88\deg$ which yielded a total eclipse (with an amplitude of 2.8 mag, when
the bright primary component is totally eclipsed, and the dominant light source
is the distant C component), making discovery easier.
However, it should be emphasized that this time-data 
are rough approximations only. Since we could not determine the position of the
node better than $\pm2\deg$, and for exact calculations one needs an 
accuracy better by one order of magnitude, these 
numbers should be refined in the future.

Considering the scientific era, one can see that our result suggests an
inclination variation of $\Delta i\approx-1\fdg6$ in the last century which
is in accordance with the statement of \citet{soderhjelm80}.
\acknowledgments

{\tiny The CHARA Array is funded by the National Science Foundation through NSF
grants AST-0307562 and AST-06006958 and by Georgia State University through the College of Arts
and Sciences and the Office of the Vice President for Research. 

e-VLBI developments in Europe are supported by the EC DG-INFSO  funded
Communication Network Developments project 'EXPReS', Contract No. 02662. 
The European VLBI Network 
is a joint facility of European, Chinese, South African and other radio 
astronomy institutes funded by their national research councils.

Zs.P. acknowledges support from the Hungarian Scientific Research Fund
(OTKA, grant no. K72515).

This research has made use of the SIMBAD database, operated at CDS, Strasbourg, 
France and has made use of NASA's Astrophysics Data System.}




\end{document}